\documentclass{elsart}

\usepackage{amssymb}
\usepackage{graphicx}

\def\BibTeX{{\rm B\kern-.05em{\sc i\kern-.025em b}\kern-.08em
    T\kern-.1667em\lower.7ex\hbox{E}\kern-.125emX}}

\begin{document}

\begin{frontmatter}

\title{Recurrence network measures for hypothesis testing using surrogate data: application to black hole light curves}

\author[label1]{Rinku Jacob}
\ead{$rinku.jacob.vallanat@gmail.com$}
\author[label1]{K. P. Harikrishnan\corauthref{cor1}}
\ead{$kp.hk05@gmail.com$}
\author[label2]{R. Misra}
\ead{$rmisra@iucaa.in$}
\author[label3]{G. Ambika}
\ead{$g.ambika@iiserpune.ac.in$}

\corauth[cor1]{Corresponding author: Address: Department of Physics, The Cochin College,  Cochin-682002, India; Phone No.0484-22224954;  Fax No: 91-22224954.} 

\address[label1]{Department of Physics, The Cochin College, Cochin-682002, India}
\address[label2]{Inter University Centre for Astronomy and Astrophysics, Pune-411007, India}
\address[label3]{Indian Institute of Science Education and Research, Pune-411008, India}

\begin{abstract}
Recurrence networks and the associated statistical measures have become important tools in the 
analysis of time series data. In this work, we test how effective the recurrence network measures 
are in analyzing real world data involving two main types of noise, white noise and colored noise. 
We use two prominent network measures as discriminating statistic for hypothesis testing using 
surrogate data for a specific null hypothesis that the data is derived from a linear stochastic 
process. We show that the characteristic path length is especially efficient as a discriminating 
measure with the conclusions reasonably accurate even with limited number of data points in the time 
series. We also highlight an additional advantage of the network approach in identifying the 
dimensionality of the system underlying the time series through a convergence measure derived from 
the probability distribution of the local clustering coefficients. As examples of real world data, 
we use the light curves from a prominent black hole system and show that a combined analysis using 
three primary network measures can provide vital information regarding the nature of temporal 
variability of light curves from different spectroscopic classes.    
\end{abstract}


\end{frontmatter}

\section{Introduction}
Detecting deterministic nonlinearity in real world data contaminated by different types of noise is a highly 
nontrivial problem. It is still one of the major challenges in nonlinear time series analysis \cite {kan}, 
though several methods and measures have been suggested over the years to address this long standing 
issue \cite {spr,sch1}. A generally accepted procedure to detect any nontrivial behavior in a time series 
is the method of surrogate data \cite {the}, for a statistical hypothesis testing, though there are 
other ways reported in literature to probe nonlinearity of time series without employing surrogates, 
under certain conditions. Examples are methods related to time-directed network properties of visibility 
graphs for testing the time-reversal asymmetry \cite {jfd,lla}. The method of surrogate data involves 
generating an ensemble of surrogates from the data. A specific null hypothesis is assumed for the data 
that there is no nontrivial character associated with it. The data and the surrogates are then subjected to 
the same analysis sensitive to this nonlinear measure. One then tries to statistically reject the null 
hypothesis for the data by comparing the results for the data and the surrogates \cite {the}, with certain 
confidence level. 

In the present analysis, we assume a specific null hypothesis that the data is generated from a linear 
stochastic process and no nonlinearity is associated with it. We generate a set of surrogate data which 
are compatible with the null hypothesis of a linear stochastic process. We then use certain measures 
derived by transforming the time series to a complex network as discriminating measures (as explained below) 
and try to reject the null hypothesis for the data. 

Though the method of surrogate analysis is very popular, there are also many challenges associated 
with it \cite {kug}. For example, generation of proper surrogate data is very important for the success of 
hypothesis testing. The method to generate surrogate data was initially introduced by Theiler et al. 
\cite {the} with the Amplitude Adjusted Fourier Transform (AAFT) surrogates. These surrogates are 
capable of testing the null hypothesis that the data come from linear as well as nonlinear static 
transformation of a linear stochastic process. An improved version of the AAFT algorithm has been suggested by 
Schreiber and Schmitz \cite {sch2,sch3} using an iterative scheme called the IAAFT surrogates, which is 
reported to be more consistent to test null hypothesis \cite {kug}. Recently, 
Nakamura et al. \cite {nak} have proposed a surrogate generation method called Truncated Fourier 
Transform (TFT) \cite {luc}. However, the surrogate data generated by this method are influenced by a 
cut-off frequency. In addition, there are also some other types of surrogate data testing reported in 
the literature, such as, cycle shuffle surrogates \cite {nakam1}, surrogates for testing pseudoperiodic 
time series \cite {nakam2} and even recurrence based surrogates \cite {thiel}, with each scheme found useful 
in particular contexts. 
In this work, we apply the IAAFT scheme to generate surrogate data using the 
TISEAN package \cite {heg}.  

The second major factor in the surrogate analysis is the choice of a discriminating measure that is 
sensitive to the nonlinearities associated with the data. In many cases,  
the correlation dimension $D_2$ and the correlation entropy $K_2$ have been used as the 
discriminating measures \cite {sprt} 
as they can be directly computed from the time series by the delay embedding method \cite {gra}. However, 
the number of data points should be sufficiently large for a proper computation of these measures. In the 
paper by Theiler et al. \cite {the}, a time reversal asymmetric statistic was introduced which required 
relatively short time series for computation. In this  
paper, we consider the use of recurrence network (RN) measures for hypothesis testing, under varying 
conditions of noise. One obvious advantage of these measures is that they can be computed with 
reasonable accuracy even when the time series is short (say $< 5000$ data points) \cite {sub1}.  
Recently, Subramaniyam et al. \cite {sub1,sub2} 
have used the RN measures for the analysis of EEG data and have shown that these measures can provide 
insights into the structural properties of EEG in normal and pathological states. Very recently, 
we have shown that the RN measures can characterize the structural changes in a chaotic attractor 
contaminated by white and colored noise \cite {rj1}.  Here our aim is to highlight 
their effectiveness as a tool for hypothesis testing in noisy environment, especially when colored noise 
is involved. We specifically show that the characteristic path length is very useful in this regard. 
Moreover, we also present a unique advantage of network based measures for analysis in that 
the degree distribution as well as the distribution of the local clustering coefficient 
of the RN provides important information regarding the dimension or the number of 
variables required to model the underlying system. We specifically derive a convergence factor using 
the standard Kullback-Leibler measure to identify the dimension beyond which the distributions tend to converge. 
Details regarding the construction of the RN and the various network measures used in this paper are 
discussed in the next section.

We use a time series from the Lorenz attractor as prototype to illustrate the effectiveness of 
using RN measures as discriminating statistic. We add different percentages of 
white and colored noise to the standard Lorenz attractor time series and the surrogates to get a 
quantitative estimate of how 
much noise can swamp the inherent nonlinear behavior and how to fix the threshold of the statistical 
measure to discriminate between nonlinearity and noise. As examples of real world data involving colored 
noise, we analyse light curves from the prominent black hole system GRS1915+105. This black hole system 
is considered to be unique with the light curves falling into $12$ spectroscopic classes \cite {bel}, whose 
details are discussed in \S 4. The system appears to randomly flip in X-ray intensity variations and 
these observed intensity variations averaged over all energy bands are grouped into $12$ different states. 
Earlier analysis \cite {kph1} using the measures $D_2$ and 
$K_2$ has strongly indicated deterministic nonlinearity for light curves in $5$ of the $12$ classes. 
Here we show that 
analysis using the network measures can provide more exact information regarding the dimensionality of the 
underlying system as well as the nature of noise contamination in different states.

Finally, it is also important to share some thoughts as to why the proposed method based on RN works. 
From a conceptual point of view, Donges et al. \cite {dong1} have shown that all RN properties can be 
analytically derived from the system's invariant density. In that case, if we generate IAAFT surrogates 
from a univariate time series of a nonlinear deterministic system, then by definition, the surrogates 
will leave the probability distribution invariant. However, the particular phase relationship between 
different parts of the reconstructed attractor changes. Hence it is important to clarify that the 
success of the proposed method is based on taking IAAFT surrogates from univariate time series and then 
use embedding of this surrogate time series, instead of taking the original multivariate time series 
and their IAAFT surrogates. The latter possibility may result in some distinctly different behavior. 

Our paper is organized as follows: In the next section, we discuss the details regarding the construction of the 
RN and the computation of the network measures to be used for hypothesis testing. In \S 3, we do surrogate 
analysis on synthetic data from the Lorenz attractor  as well as data obtained by adding different percentages 
of white and colored noise to the Lorenz data. Analysis of the real world data from the black hole system is 
presented in \S 4. Conclusions are drawn in \S 5.

\section{Recurrence networks and related measures}
Recurrence is a fundamental property of every dynamical system \cite {eck}. This property has been used for developing 
a two dimensional visualization tool called the recurrence plot (RP) \cite {mar1}. Considerable information 
regarding the nature of the underlying dynamical system can be obtained from the RP. To generate the RP, one has 
to first reconstruct the attractor from the time series using the delay embedding method \cite {gra}. The recurrence 
of trajectory points can then be represented by a two dimensional plot by putting a ``dot'' at the 
$\imath, \jmath^{th}$ element if the metric distance between the two points $\imath$ and $\jmath$ on the 
reconstructed attractor is 
less than or equal to a threshold value denoted by $\epsilon$. There will obviously be a diagonal line in 
the RP representing the recurrence of each point with itself. Quantification of the dynamical properties can be 
obtained from the RP using the so called recurrence quantification analysis \cite {zbi}. 

Recently, due to the success of the complex network theory in various fields \cite {str,new1}, time series analysis 
based on the statistical measures of complex networks has also gained a lot of attention. A specific 
advantage of this approach is that it can be applied to short and nonstationary time series data \cite {don1}. 
For this, the time series is first converted to a complex network defined in an abstract space with a set of 
nodes $\mathcal N = {1,2,3,...,N}$ connected by a set of links between 
the nodes. Several methods have been suggested  \cite {sma,xu} to transform a time series to a 
complex network. Here we consider the method of $\epsilon$ - RNs \cite {dong1,don2} which is based on the 
property of recurrence of a dynamical system and is closely related to the RP defined above. 

\begin{figure}
\begin{center}
\includegraphics*[width=16cm]{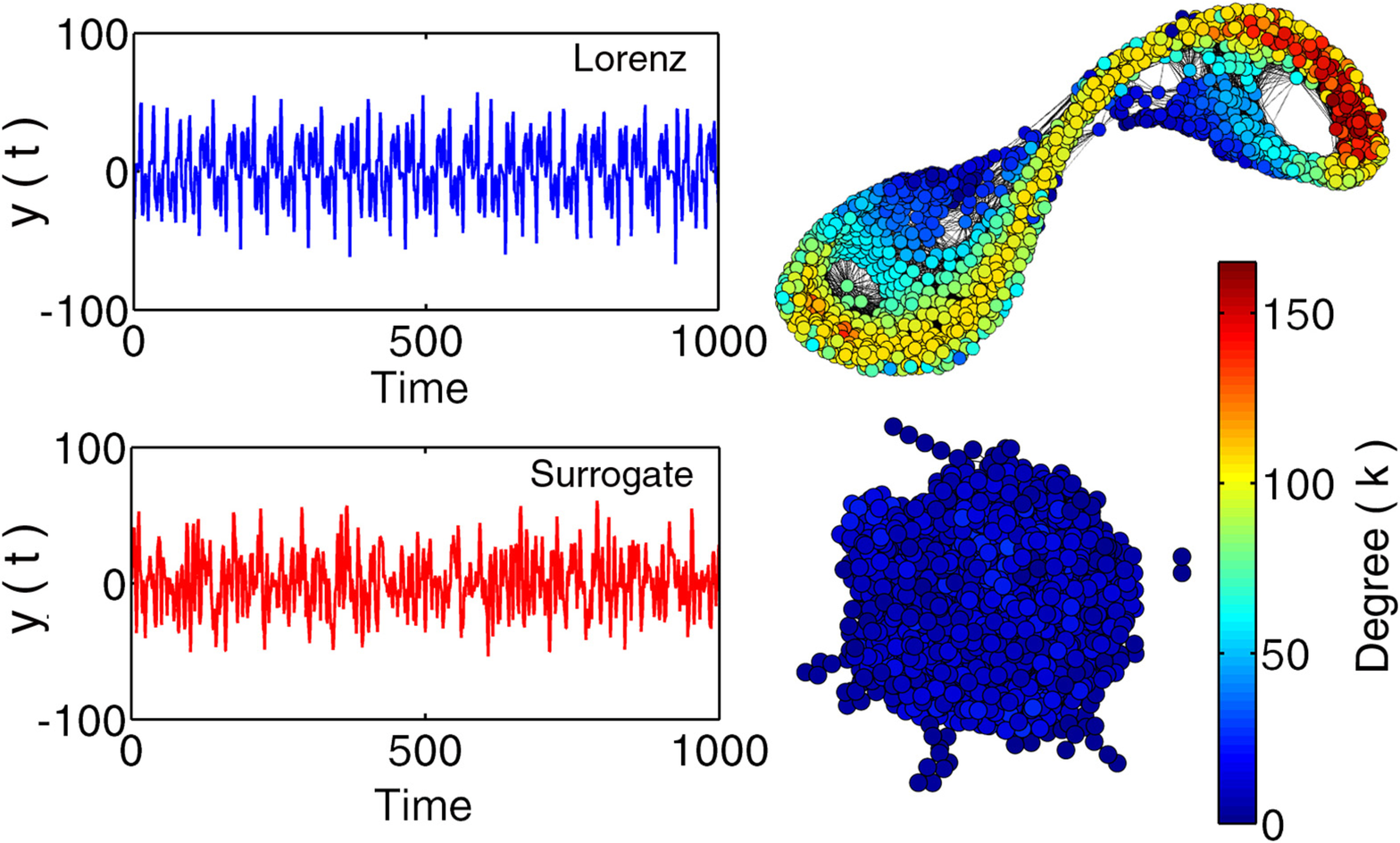}
\end{center}
\caption{Time series from the standard Lorenz attractor generated from Eq.~\ref{eq:4} and the RN constructed 
from the time series are shown in the top panel. The surrogate time series and the corresponding RN are 
shown in the bottom panel. Color code represents the variation of the node degree as indicated.} 
\label{f.1}
\end{figure}

\begin{figure}
\begin{center}
\includegraphics*[width=16cm]{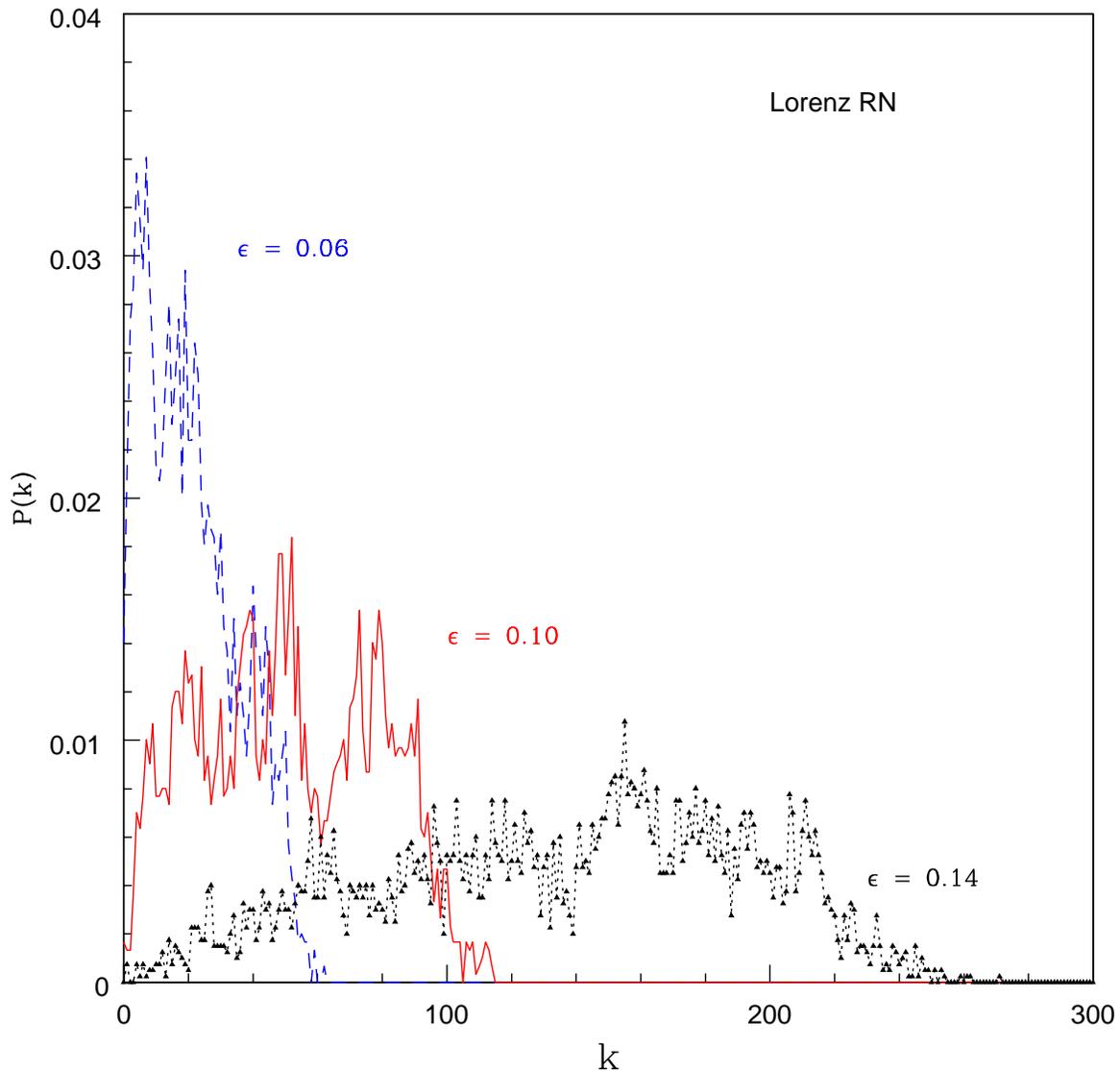}
\end{center}
\caption{Degree distribution for RNs constructed from the Lorenz attractor using three values of 
$\epsilon$ as indicated in the figure, with $M = 3, N = 3000$. The change in the degree distribution 
is an indicator for selecting the proper value of recurrence threshold (see text).} 
\label{f.2}
\end{figure}

The attractor is first reconstructed from the given time series using a suitable time delay $\tau$. 
Here we  use the automated algorithmic scheme proposed by us \cite {kph2} for attractor reconstruction. 
The scheme involves transforming the time series into a uniform deviate so that the embedded attractor 
always remains in a unit cube of dimension $M$. The time lag $\tau$ for embedding is chosen as the 
value of the de-correlation time. A recurrence matrix $\mathcal R$ is 
constructed by a method identical to the construction of the RP. If the distance between two points 
$\imath$ and $\jmath$ on the attractor is $\leq \epsilon$, the recurrence threshold, the corresponding 
element on the matrix is $1$ and otherwise $0$. Two nodes $\imath$ and $\jmath$ are considered to be connected 
only if the corresponding matrix element in $\mathcal R$ is $1$. The adjacency matrix $\mathcal A$ of the 
complex network is obtained by removing the self loop (diagonal elements) from the matrix $\mathcal R$. 
It is obvious that the matrix $\mathcal A$ is a binary, symmetric matrix implying that the resulting 
complex network is unweighted and undirected called the RN.

The number of nodes connected to a reference node is called its degree denoted by $k$ and if 
$N$ is the total number of nodes, $k/N$ is called the degree density. Note that any node in the RN is 
connected to nodes in its neighbourhood only since the range of connection cannot exceed the threshold value 
$\epsilon$. In other words, only short range connections exist in a RN. Due to this, the probability 
density variations over the attractor is mapped onto the local degree density variations over the 
network as the attractor is transformed into a complex network. Consequently, the node structure of 
the complex network closely follows the invariant density of the embedded attractor \cite {dong2}.   
On the other hand, the topology of the network is reflected in the degree distribution denoted by $P(k)$ 
which is a probability distribution representing how many nodes have a given degree $k$. 

Though the method of transforming the time series to a complex network outlined above is simple and 
straight forward, care must be taken to ensure that the resulting RN can characterize the properties 
of the embedded attractor. This is achieved by the proper choice of the recurrence threshold 
$\epsilon$ and the embedding dimension $M$. We have recently shown \cite {rj2} that the value of 
$\epsilon$ is closely related to the choice of $M$. We have also proposed a general method for 
the construction of RN from a time series, which we follow here. The basic criterion that we use to 
select the recurrence threshold $\epsilon$ is that the resulting RN should have a giant component  
and the $\epsilon$ value obtained using this criterion is approximately 
the same for different time series for a given $M$, due to the uniform deviate transformation. For 
$M = 3$, the value of $\epsilon$ obtained is $0.1$. As an illustrative example, we show 
in  Fig.~\ref{f.1}, the time series from the standard Lorenz attractor and its RN (constructed with 
$M = 3$ and $\epsilon = 0.1$) along with a surrogate 
of the time series and the corresponding RN. We use the $y$ - component of the Lorenz system to generate  
the time series with a time step of $0.05$. 

Apart from the presence of the giant component, another method to choose the value of  
$\epsilon$ is to look at the degree distribution of the resulting RN. When $\epsilon$ 
is less than the threshold value, there will be large number of nodes with $k \sim 0$ 
(unconnected nodes) and the 
resulting $k$ values in the network will be within a small range, centered around a small average 
value $<k>$. On the other hand, if $\epsilon$ is much greater than the optimum threshold, the network is 
over connected which may not be easily identified looking at the network. However, in the degree 
distribution, $P(k)$ will exhibit small values for a wide range of $k$, which 
implies that the structure of the attractor has not been properly captured by the RN. This will also make 
the characteristic path length of the network tending to a much smaller value. In between these two extremes, for a 
small range of $\epsilon$, the resulting network becomes a proper representation of the attractor. 
This is shown in Fig.~\ref{f.2} for the Lorenz attractor, where degree distribution for RNs with 3 
values of $\epsilon$ are shown, using $M = 3$ in all cases.

Once the time series is transformed into a complex network, the structural properties of the 
reconstructed attractor can be characterized by the statistical measures of the complex network. Though  
many different statistical measures can be defined for a complex network \cite {new2}, here we concentrate on 
two for surrogate analysis: an averaged local measure called the 
average clustering coefficient (CC) and a global network measure, the characteristic path length (CPL). 
The first one can be defined through a local clustering index $C_v$. For the reference node $v$, its 
value can be determined by finding how many nodes connected to $v$ are also mutually connected: 
\begin{equation}
C_v = {{\sum_{i,j} A_{vi}A_{ij}A_{jv}} \over {k_v(k_v - 1)}}
  \label{eq:1}
\end{equation} 
where $A_{vi}$ are the elements of the adjacency matrix.  
The average value of $C_v$ is taken as the CC of the whole network:
\begin{equation}
CC = {{1} \over {N}} \sum_{v} C_v
  \label{eq:2}
\end{equation} 
The CPL is defined through the shortest path length $l_{ij}^s$ between any pair of nodes $(\imath,\jmath)$ 
in the network. It represents the minimum number of nodes to be covered to reach from a reference 
node $\imath$ to any other node $\jmath$ in the network. To calculate CPL, we first compute $l_{ij}^s$ for 
all the nodes in the network and then the global average is found:
\begin{equation}
CPL = {{{1} \over {N}} \sum_{i=1}^N ({{1} \over {N-1}} \sum_{i \neq j=1}^{N-1} l_{ij}^s)}
\label{eq:3}
\end{equation} 
Here we try both CPL and CC as discriminating measures for hypothesis testing as discussed in the 
next section. Apart from these two, we also use the degree distribution $P(k)$ and the distribution of 
the local clustering coefficient $P(C_v)$ to gain information 
regarding the dimension of the underlying system as explained in the next section. 

\section{Analysis of synthetic data}
We first analyze data from the standard Lorenz attractor whose equations are given by:
\begin{eqnarray}
    {{dx} \over {dt}} & = & \sigma(y - x)     \nonumber  \\
    {{dy} \over {dt}} & = & x(r - z)    \nonumber  \\
    {{dz} \over {dt}} & = & xy - bz    
    \label{eq:4}
\end{eqnarray} 
with the parameter values $\sigma = 10$, $r = 28$ and $b = 8/3$.  
As mentioned above, one advantage of using network measures is that 
they can be effectively computed from a relatively low number of nodes in the network. 
Specifically, since the number of data points in all the black hole light curves is 
$\sim 3000$, we use the same number of data points for the Lorenz attractor time series as well. 
To study the effect of noise, we generate an ensemble of data sets by adding different percentages of 
white and colored noise to the Lorenz data. The colored noise essentially produces a random fractal 
curve with power varying as $p(f) \propto 1/f^s$, where $s$ can, in practice, vary from $1$ to $2$. 
However, the two most prominent cases in terms of occurence in real world are $s = 1$, called 
$1/f$ noise (which mimics the Brownian motion) and $s = 2$, called the red noise. We have generated 
both and done the analysis adding to the Lorenz data. While the results for $s = 1$ is very close to 
that of white noise, those for $s = 2$ are very different and hence we use this case to represent the 
colored noise in general with $s$ varying from $1.5$ to $2$. 
Surrogate analysis is performed with $20$ surrogates for each data generated by the 
TISEAN package \cite {heg}.

\begin{figure}
\begin{center}
\includegraphics*[width=16cm]{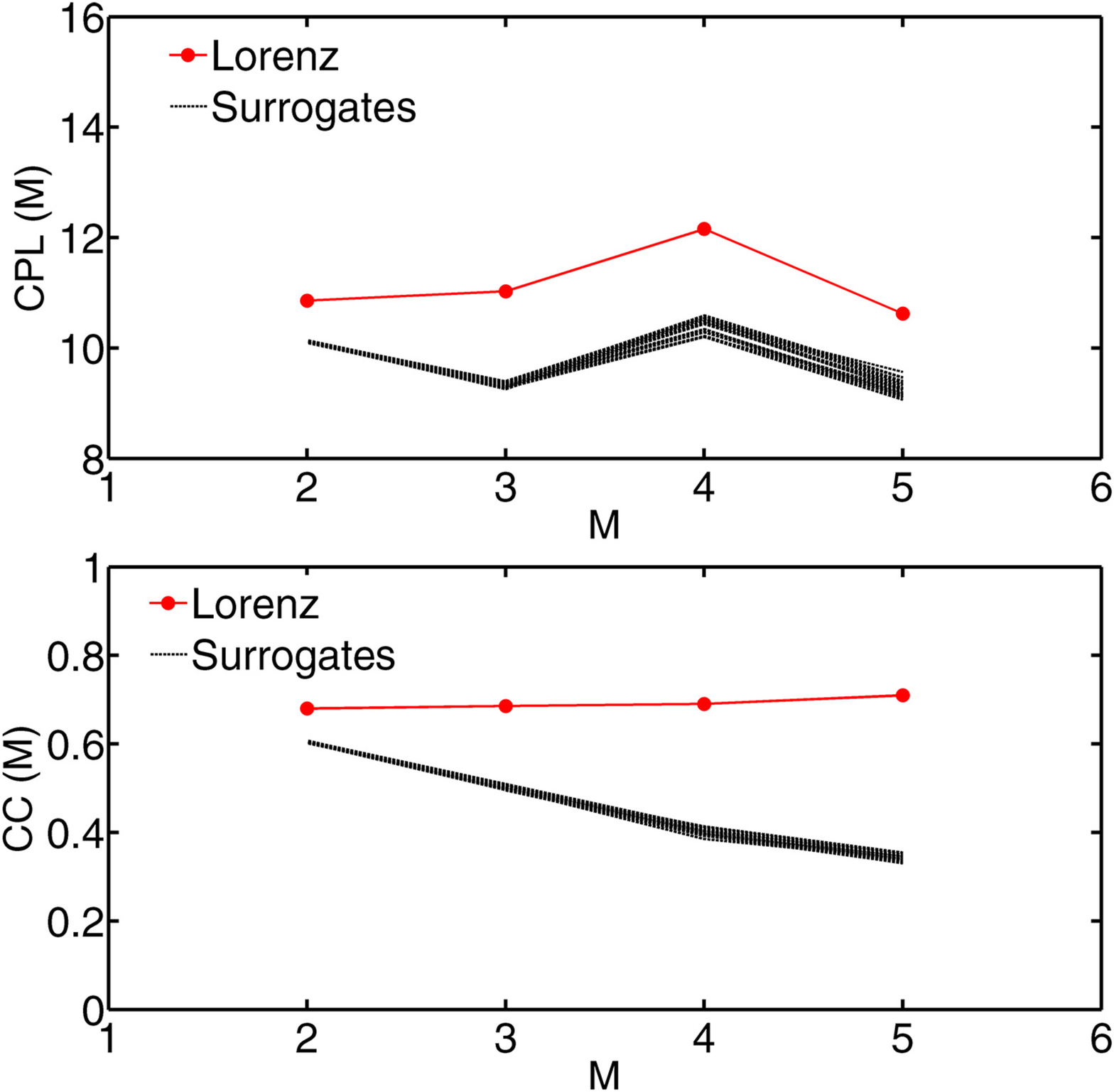}
\end{center}
\caption{Results of RN analysis of the Lorenz attractor time series with $20$ 
surrogates using both CPL and CC as quantifying measures. Results for surrogates are shown as dotted  
black lines. For each $M$, the corresponding value of recurrence threshold $\epsilon$ is used for 
computing CPL and CC.} 
\label{f.3}
\end{figure}

\begin{figure}
\begin{center}
\includegraphics*[width=16cm]{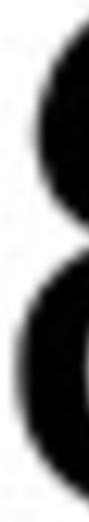}
\end{center}
\caption{RN analysis of Lorenz data added with two different percentages of white noise and their 
surrogates with CPL and CC as quantifying measures.} 
\label{f.4}
\end{figure}

\begin{figure}
\begin{center}
\includegraphics*[width=16cm]{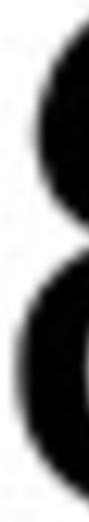}
\end{center}
\caption{Same as Fig.~\ref{f.4}, but with additive red noise instead of white noise.} 
\label{f.5}
\end{figure}

In Fig.~\ref{f.3}, we show the result of RN analysis of the  Lorenz attractor time series and its 
surrogates using both CPL and CC as quantifying measures. For each $M$, we use the value of $\epsilon$ 
that satisfies the primary criterion of the existence of a 
giant component in the RN for computing the network measures, as 
discussed in detail in our scheme \cite {rj2}. It is clear that the null hypothesis that the data comes 
from a linear stochastic process can be 
rejected with high confidence level in both cases. The question may naturally arise why the considered 
linear surrogates lead to different RN properties than the original data. Though the AAFT surrogates 
keep the distribution of time series conserved, the nonlinear structures present in the data are 
destroyed in the surrogates leading to  different values of measures after embedding. Effectively, 
this leads to an altogether 
different RN compared to that from data, as is evident from Fig.~\ref{f.1}. 

In order to quantify the results, we use a 
statistical measure proposed earlier \cite {kph1}, namely, 
\emph{the normalised mean sigma deviation or nmsd}. For CC, the measure can be defined as 
\begin{equation}
nmsd^2 = {\frac{1}{M_{max} -1} \sum_{M = 2}^{M_{max}} \large (\frac{CC (M) - < CC^{surr} (M) >}{\sigma^{surr}_{SD} (M) }\large )^2} 
  \label{eq:5}
\end{equation}  
with a similar expression for CPL. Here $M_{max}$ is the maximum embedding dimension for which the analysis
is undertaken, $CC(M)$ is the CC for the data, $<CC^{surr} (M)>$ is the average CC for $20$ 
realizations of the surrogate data  and $\sigma^{surr}_{SD} (M)$ is the standard deviation of $CC^{surr} (M)$. 
For the Lorenz time series, $nmsd(CC) = 38.46$ and $nmsd(CPL) = 26.37$. For pure white noise, the 
$nmsd$ values are found to be $nmsd(CC) = 1.92$ and $nmsd(CPL) = 1.64$ while the corresponding values for 
red noise are obtained as $nmsd(CC) = 2.26$ and $nmsd(CPL) = 2.08$. The value of $nmsd$ can be used 
effectively as a quantitative measure to reject the null hypothesis.

The RN analysis of data and surrogates is performed by adding different percentage of white and colored noise to the 
Lorenz data. The results for two noise levels $5\%$ (SNR $20$) and $20\%$ (SNR $5$) are shown in Fig.~\ref{f.4} 
for white noise and in Fig.~\ref{f.5} for red noise. From Fig.~\ref{f.4}, we find that the data and 
surrogates become hardly distinguishable when the noise level reaches $20\%$ for both CPL and CC. On the other 
hand, from Fig.~\ref{f.5} with additive red noise, this is true only for CPL (left panel) and not for CC. 
Moreover, two other results can also be inferred from these figures: 

i) The values of data and surrogates show much deviation beyond the actual dimension of the system. 

ii) The CPL of pure data are much above the surrogates. As the noise level increases, the value of 
the measure for the original 
data systematically approaches that of the surrogates. However, in some cases, the CPL of data can be below that of 
surrogates, but the difference decreases with increase in noise. 
   
\begin{figure}
\begin{center}
\includegraphics*[width=16cm]{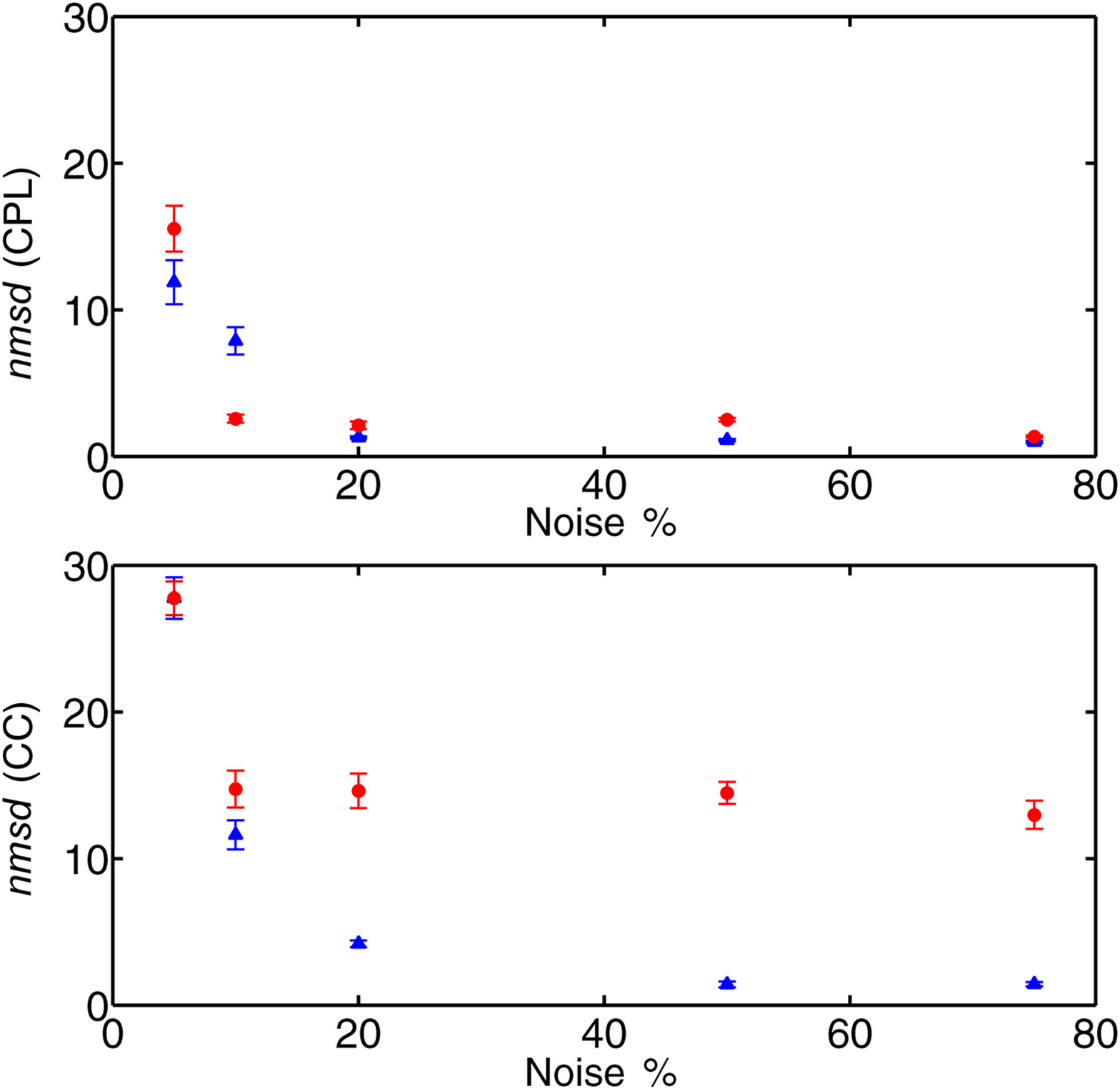}
\end{center}
\caption{Variation of $nmsd$ (see text) for both CPL and CC with $\%$ of noise added to 
Lorenz data. Solid triangles are for white noise and solid circles are for additive red noise.} 
\label{f.6}
\end{figure}

To get a statistically more relevant result, we generate Lorenz data with $10$ different initial conditions and 
add noise on each so that we have $10$ different data sets for each level of added noise. By performing the 
surrogate analysis, we compute the $nmsd$ for each percentage with an error bar obtained from the standard 
deviation. The variation of $nmsd(CC)$ and $nmsd(CPL)$ with $\%$ of noise for both white and red noise is 
shown in Fig.~\ref{f.6}. 

From the figure it becomes clear that the CC of the RN is not much affected by red noise contamination as the 
$nmsd(CC)$ remains high even with high percentage of red noise.  In other words, in terms of CC, it is difficult to 
distinguish between a low dimensional chaotic attractor and red noise by our scheme. This implies that for RN 
analysis of data and surrogates where red noise is expected, such as astrophysical light curves, CC is not a good 
discriminating measure. On the other hand, CPL seems to be sensitive to both white and colored noise 
contamination and can be used as an effective discriminating statistic between data and surrogates. Hence, 
in the analysis of the 
black hole light curves below, we use only CPL as the discriminating measure. 
We now fix a lower limit for the value of $nmsd$ for rejecting the null hypothesis based on our results. 

From Fig.~\ref{f.6}, we find that as the noise level reaches $20\%$, the $nmsd(CPL)$ for both white and 
colored noise becomes very small ($< 2.0$). Hence we fix a conservative limit of $10\%$ for noise level 
above which detection of nontrivial structures in the data is considered to be difficult. The average 
value of $nmsd$ corresponding to this noise level, namely $5.0$, is fixed as the threshold  for 
rejecting the null hypothesis. It could be a point of argument whether one can fix a common threshold 
for \emph {nmsd} applicable to all the different types of systems. This is because, sensitivity of 
noise can be system-specific and a good threshold $nmsd$ for one may not be so for another. Though this is 
true in principle, we find that the proposed threshold works in practice for a variety of systems. 

In the case of synthetic time series, such as the one from the Lorenz attractor considered above, the 
number of variables involved and hence the embedding dimension $M$ are known \emph{a priori}. However, 
this information is absent in the case of observational data. Just like other measures, such as $D_2$ and 
$K_2$, the network 
measures are also likely to be inaccurate if the applied embedding dimension is less than the actual 
dimension of the system. Two methods are commonly used to select the proper embedding dimension $M$. 
The more popular method is the one based on false nearest neighbor \cite {kenn}. Here one looks at the 
changes in the nearest neighbors to a reference point as the dimension increases from 
$M \rightarrow M+1$. When the attractor is unfolded completely, change in nearest neighbors $\rightarrow 0$. 
The second method is to check for a saturated value of $D_2$ and then take the next higher integer value 
as $M$. But both these methods become difficult if the data is short and noisy. Here we show that in 
such situation, a network based measure can be used to find an appropriate value of $M$. 

\begin{figure}
\begin{center}
\includegraphics*[width=16cm]{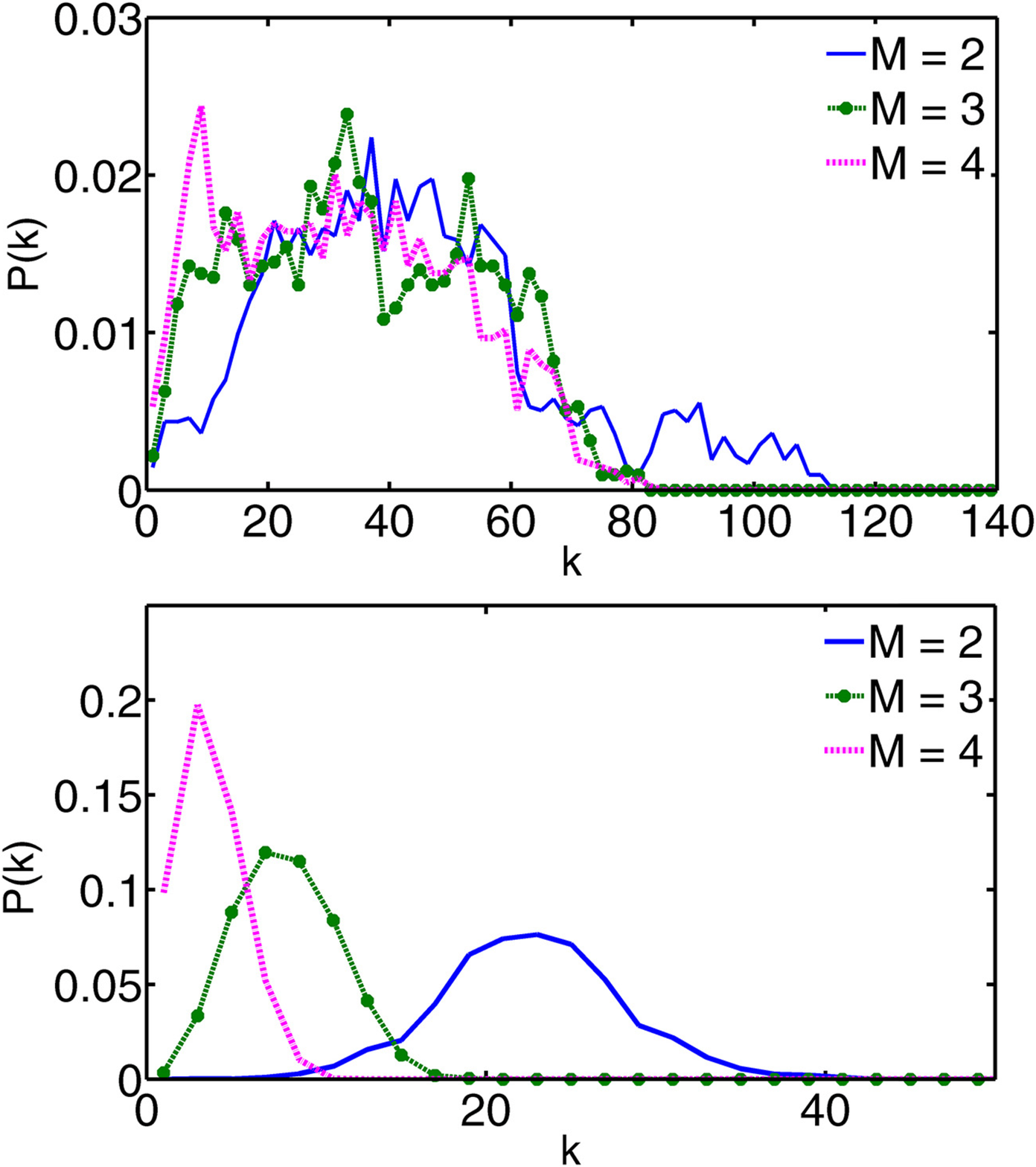}
\end{center}
\caption{(Top) Degree distribution of the RN constructed from the Lorenz attractor 
time series with $M = 2$ (solid line), $M = 3$ (solid circles connected by dashed line) and $M = 4$ 
(dotted line). (Bottom) The same for a white noise time series.} 
\label{f.7}
\end{figure}

As shown by us recently \cite {rj1}, the degree distribution can give information 
regarding the number of variables required to characterize the underlying attractor. We find that 
for RN from chaotic time series and even colored noise, the degree distribution shows approximate 
convergence beyond the 
actual dimension of the system, a behavior identical to that of $D_2$. The reason is that the attractor 
gets confined to a sub space of the total phase space and does not change for further increase in  $M$. 
However, for data dominated by white noise, the degree distribution keeps on shifting without showing 
convergence. This result is explicitly shown in Fig.~\ref{f.7} taking time series from the Lorenz attractor 
(upper panel) and white noise (lower panel) and computing the degree distribution for 
various $M$ values. It is also known \cite {rj2} that the variation of the degree distribution 
with $M$ for white noise is similar to that of the surrogate data.

\begin{figure}
\begin{center}
\includegraphics*[width=12cm]{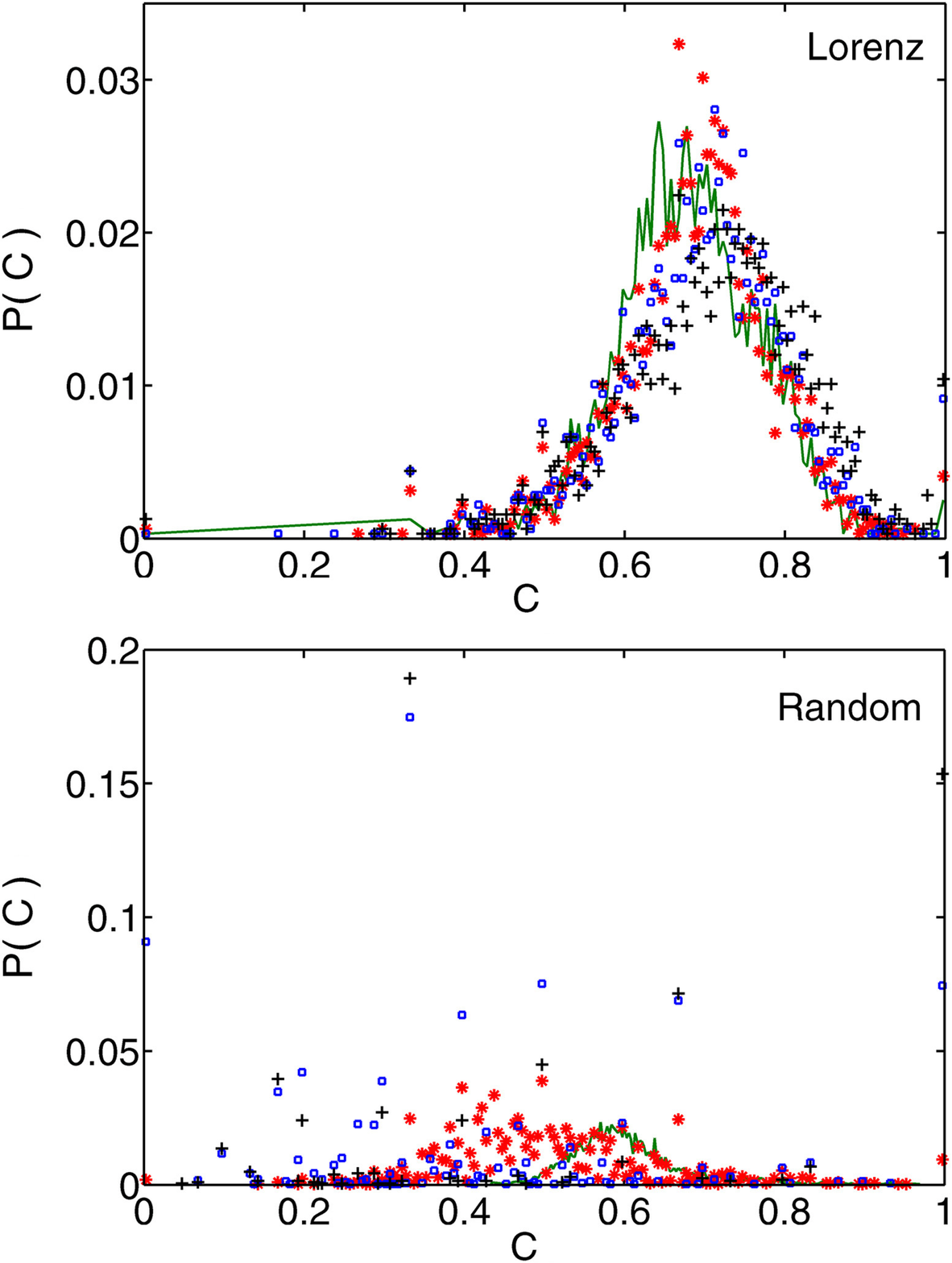}
\end{center}
\caption{Probability distribution of the local clustering coefficient $P(C)$ for RNs from the Lorenz 
attractor time series (top panel) and pure white noise (bottom panel). In both cases, the results 
are shown for $M = 2$ (green solid line), $M = 3$ (red star points), $M = 4$ (blue open squares) and 
$M = 5$ (black crosses). While the distributions show approximate convergence for $M \geq 3$ in the 
top panel, the points are scattered for all $M$ values in the bottom panel.} 
\label{f.8}
\end{figure}

However, the above observations regarding convergence are subjective since the distributions for two 
$M$ values can never coincide exactly. Hence it is desirable to have an objective measure to 
quantify the convergence of two distributions. Such a measure has been defined in the literature 
in terms of the Kullback - Leibler (K-L) divergence \cite {kull1,kull2}. This measure was originally 
introduced in probability theory to compare and quantify the difference between two probability 
distributions $P$ and $Q$. Specifically, the K-L divergence from $Q$ to $P$, denoted by 
$D_{KL}(P|Q)$, is a measure of the amount of information lost when $Q$ is used to approximate $P$. 
For discrete probability distributions, the K-L divergence from $Q$ to $P$ is defined as \cite {mack}:  
\begin{equation}
D_{KL} (P|Q) = \sum_i P(i) \log {{P(i)} \over {Q(i)}}
  \label{eq:6}
\end{equation}
For a continuous distribution, the summation is replaced by integration. 

Though this is a useful measure, we cannot apply it directly in the case of degree distribution since 
the range of $k$ values is generally different for two degree distributions. To overcome this 
difficulty, we consider the probability distribution of the local clustering coefficient $P(C)$ over 
the entire RN. We find that, like the degree distribution, $P(C)$  
also reflects the intrinsic nature of the time series and more importantly, it is ideal to 
apply the K-L measure. To get the probability $P(C)$, we compute how many nodes in the RN have a 
given value of $C$ and normalise this with respect to the total number of nodes in the network. 
The advantage here is that $C$ always varies in the unit interval and hence the comparison between 
two distributions is straightforward. 

We first check this measure for RNs from a standard chaotic time series and white noise. For this, 
we construct the RN for $M$ values from $2$ to $6$. We do not consider $M > 6$ since the number of 
nodes in the network is only $\sim 3000$. In Fig.~\ref{f.8}, we show the $P(C)$ variation for the RNs 
from standard Lorenz attractor time series (upper panel) and random time series (lower panel) for 
$M$ values from $2$ to $5$. The difference between the upper and lower panels is obvious. For the 
Lorenz system, the distributions have a structure and show convergence for $M > 2$ whereas, the same for a 
random series is scattered throughout the unit interval without showing any convergence. We have also 
checked this for other types of noise. For red noise, the distribution is found to converge  
beyond $M = 3$, while for $1/f$ noise, the distribution 
behaves almost identical to that of white noise with no convergence in $M$. We now compute 
the K-L measure by comparing two distributions at a time for consecutive $M$ values using the equation:
\begin{equation}
 D_{KL}(P_M|P_{M+1}) = |\mu_M - \mu_{M+1}| + \sum_{C} P_M (C) \log {{P_M (C)} \over {P_{M+1} (C)}}
 \label{eq:7}
\end{equation}
where $\mu_{M}$ is the average value of $C$ for dimension $M$ and $\mu_{M+1}$ is that for dimension 
$(M+1)$. These values are added to capture the difference due to the displacement of one profile from 
the other. The calculation is repeated by taking the distributions for two successive $M$ values at a 
time changing $M$ from 2 to 6. The values obtained are shown in Table 1. If the values show convergence 
above a particular $M$ value, it is taken as the required embedding dimension. For example, 
the values clearly show convergence for the Lorenz attractor beyond 
$M=2|M=3$ while they keep on fluctuating for random time series. Thus, for the Lorenz attractor and red noise,  
$M = 3$ can be chosen as the required minimum embedding dimension from the table while no such 
dimension can be chosen for the white noise.  
This measure also provides the minimum $M$ value that should be used for 
computing the network measures from the RNs from real data whose dimension is unknown.  
We make use of this measure  combined with the results obtained from RN analysis given above to 
distinguish between white noise and colored noise contamination in a time series.   

\begin{figure}
\begin{center}
\includegraphics*[width=16cm]{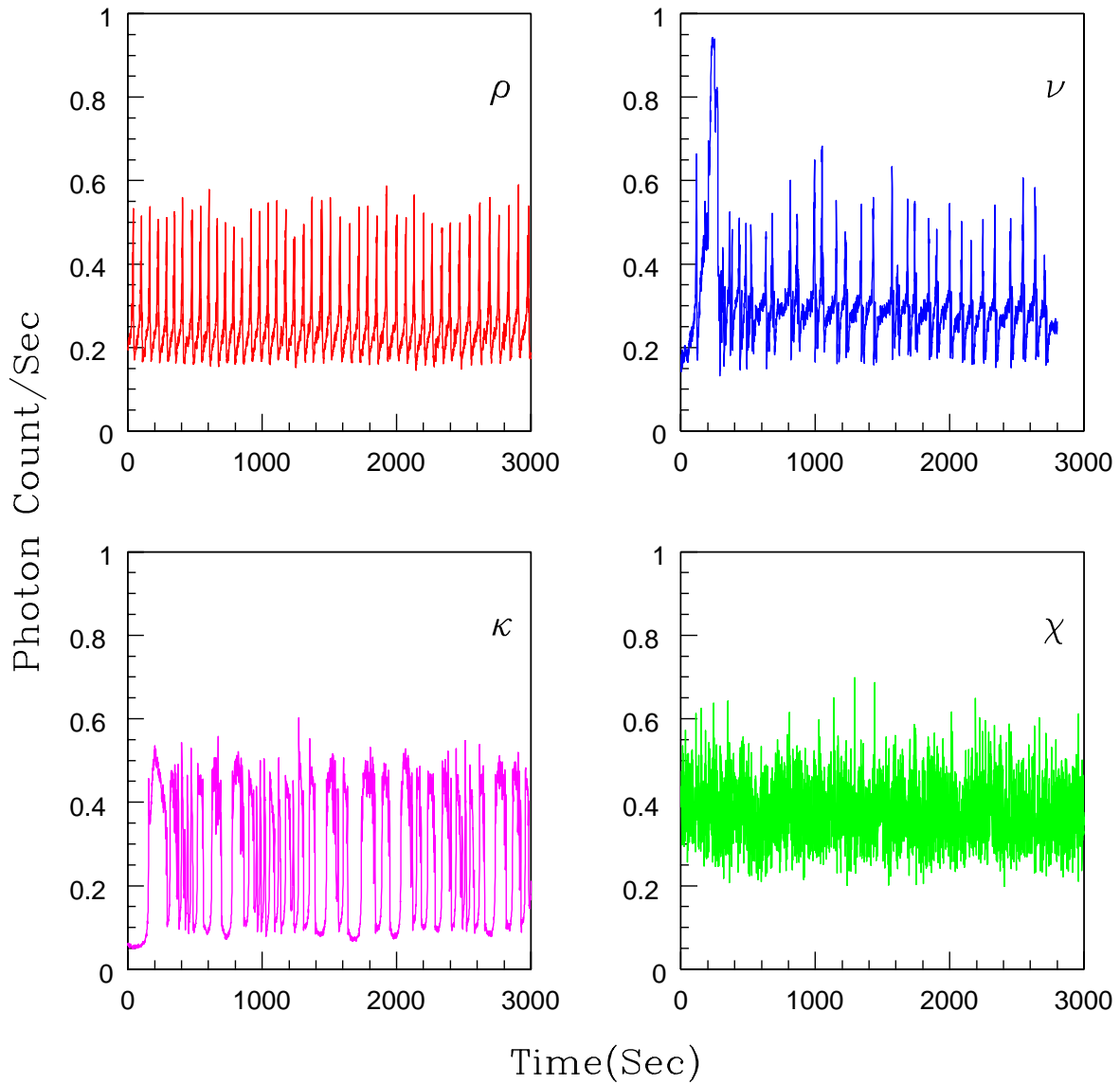}
\end{center}
\caption{Representative light curves from four spectroscopic classes of the black hole 
system GRS1915+105. The signal amplitudes of all the light curves have been rescaled into the   
unit interval. Note that the signal from $\chi$ state is much weaker compared to others.} 
\label{f.9}
\end{figure}

\begin{table}[h]
\caption{Variation of the K-L measure obtained by comparing the probability distributions of local 
CC of the RNs for successive embedding dimensions for time series from Lorenz attractor,  
white noise and the light curves from all GRS states. The last column indicates the embedding dimension  
at which the measure saturates in each case. ``NS'' indicates no saturation.}
\begin{center}
\begin{tabular}{cccccc}
\hline
\emph{System} & $D_{KL}(P_{2}|P_{3})$ & $D_{KL}(P_{3}|P_{4})$ & $D_{KL}(P_{4}|P_{5})$  & $D_{KL}(P_{5}|P_{6})$ & $M$  \\
\hline
\hline

Lorenz & 0.1962   & 0.1143    & 0.1109    & 0.1004  & 3      \\
White Noise & 0.5498   & 1.6329    & 0.8358    & 1.162 & NS    \\
Red Noise & 0.238   & 0.109    & 0.107    & 0.116 & 3    \\
&  &  &  & &    \\
$\rho$ & 0.2115   & 0.1566   & 0.0897    & 0.0834  & 4         \\
$\nu$  & 0.4687   & 0.1684     & 0.1177     & 0.1154  & 4      \\
$\beta$ & 0.8207    & 0.1445     & 0.1497    & 0.1330  & 3      \\
$\theta$ & 0.4182    & 0.1303     & 0.1255    & 0.1271  & 3     \\
$\alpha$ & 0.6375     & 0.1511    & 0.1108    & 0.1144  & 4     \\
$\kappa$ & 1.154    & 0.1918     &  0.1321   &  0.1287  & 4      \\
$\lambda$ & 0.3163    & 0.1368    & 0.1306    & 0.1427  & 3       \\
$\mu$ &  0.3853    &  0.2030    & 0.1223     & 0.1154   & 4      \\
$\gamma$ & 1.584     & 1.002     & 0.5362     & 0.8904  & NS       \\
$\delta$ & 1.254     & 0.426     & 0.714     & 0.148    & NS     \\
$\phi$ & 2.026     & 0.936     & 0.418     & 0.219      & NS   \\
$\chi$ & 2.036     & 0.875     & 1.269     & 0.137      & NS   \\

\hline
\end{tabular}
\label{tab:1}
\end{center}
\end{table}

\section{Analysis of black hole data}
In this section, we apply the network measures to analyse the X - ray light curves from the black hole 
binary GRS1915+105 to check how effective the network measures are in finding deterministic 
nonlinearity in real world data. The light curves from the black hole system have been classified 
into $12$ spectroscopic classes, labeled by $12$ different symbols ($\alpha$, $\beta$, $\rho$, $\nu$, 
$\theta$, $\kappa$, $\lambda$, $\mu$, $\delta$, $\phi$, $\gamma$ and $\chi$),  
by Belloni et al. \cite {bel} based on Rossi X-Ray Timing Explorer (RXTE) 
observation. The light curve for a given Observation ID can be obtained from the 
standard products \cite {hea}, which provide a $0.125$ second time resolution summed over all 
energy channels. It is better to have continuous data without gaps though RN analysis can be done 
on data involving gaps which is technically more challenging.  
For each class, we have extracted a few continuous 
segments for the analysis. The light curves in all cases have been generated after rebinning to a 
time resolution of 1 second (to minimise noise), resulting in continuous data of length in the range 
$3000$ to $3500$. More details regarding the data are given elsewhere \cite {rm1}. For each class, 
$6$ different light curves ($6$ observation IDs) were generated for the analysis. 

\begin{figure}
\begin{center}
\includegraphics*[width=16cm]{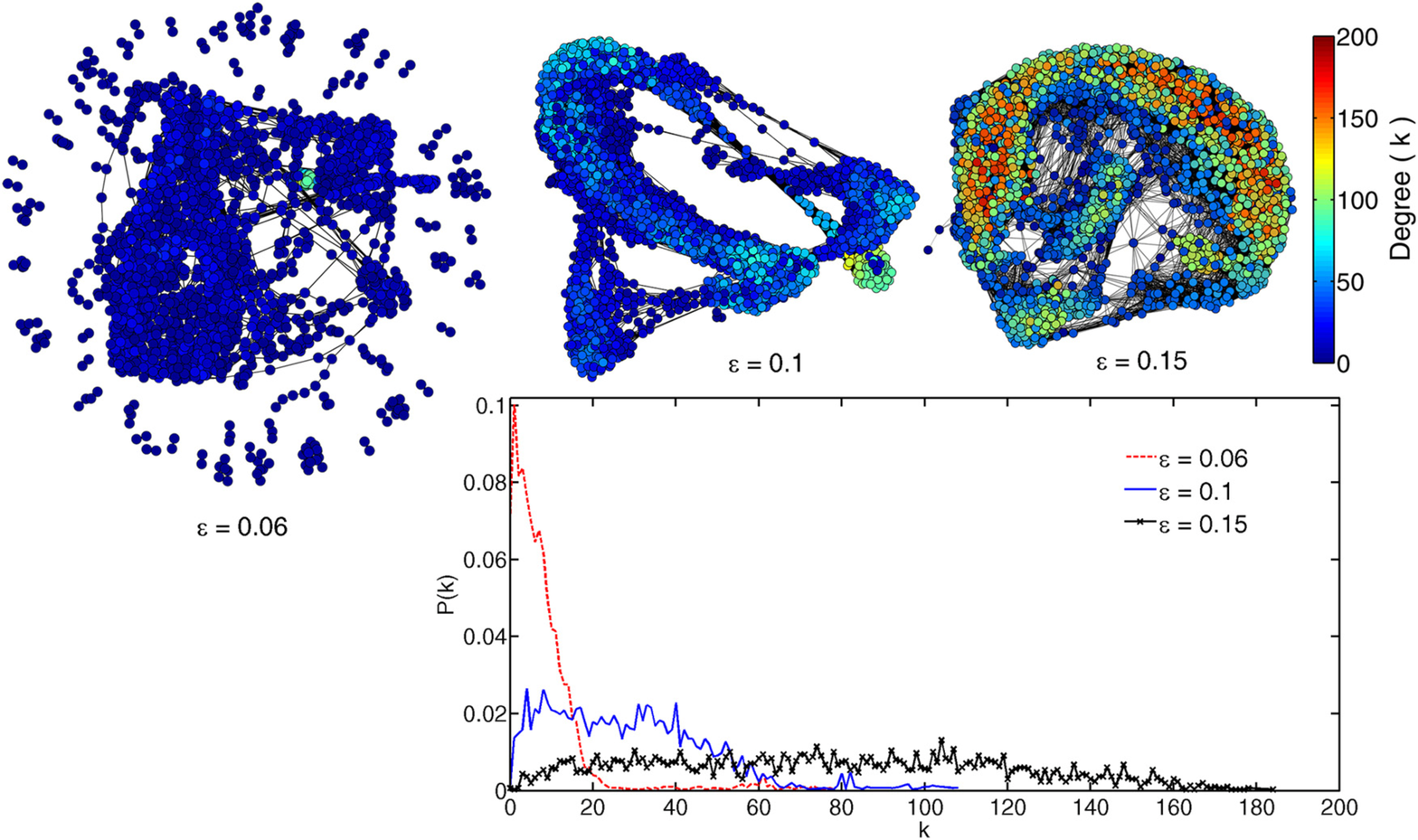}
\end{center}
\caption{RNs constructed from the light curve of $\nu$ state taking $M = 3$ and using three different 
values (as indicated) of $\epsilon$ shown in the upper panel. The degree distributions for the three 
networks are shown in the bottom panel. The networks and the distributions justify the selection of 
$\epsilon = 0.1$ as the threshold for $M = 3$. } 
\label{f.10}
\end{figure}

\begin{figure}
\begin{center}
\includegraphics*[width=16cm]{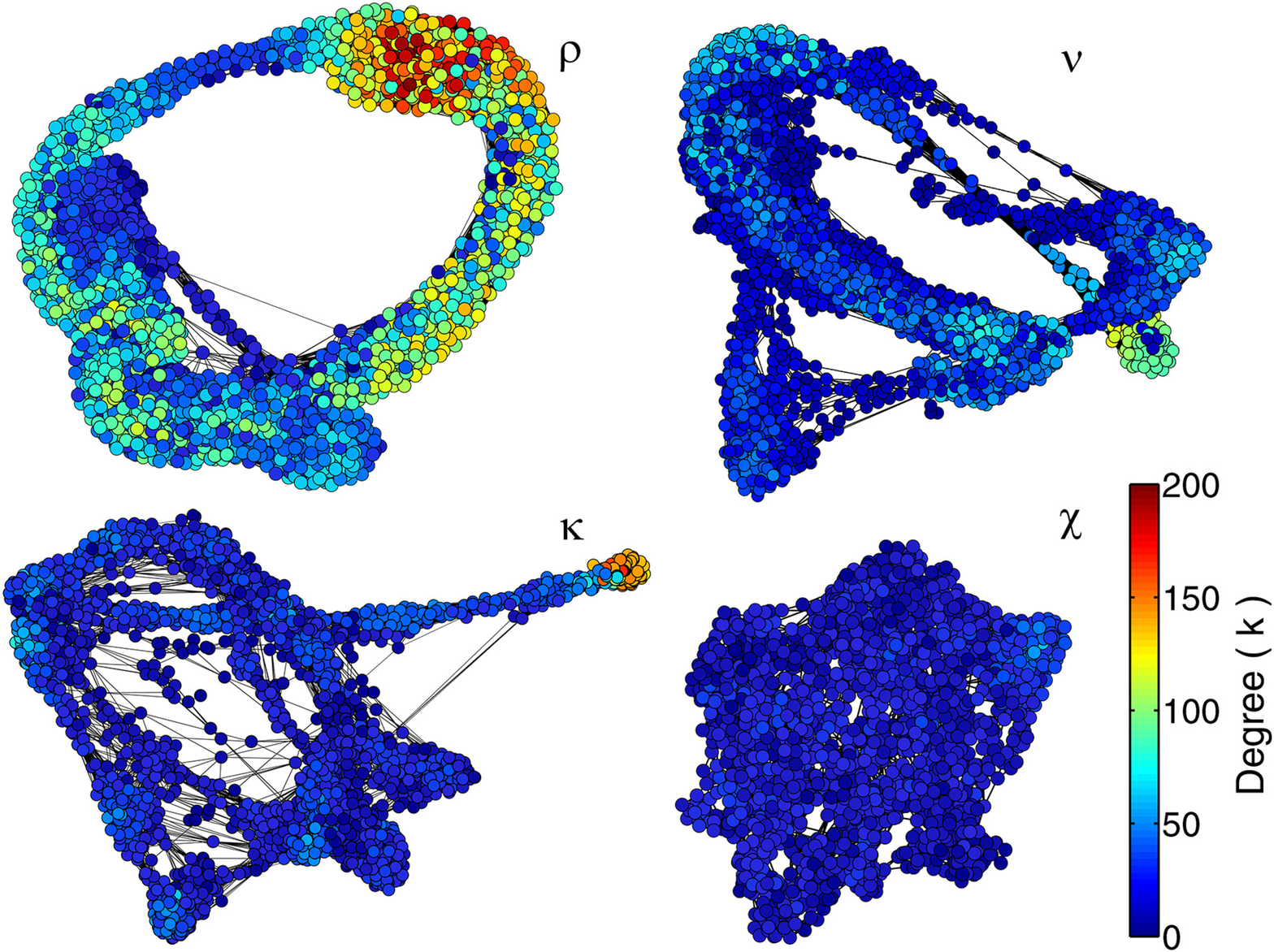}
\end{center}
\caption{RNs constructed from the four light curves shown in Fig.~\ref{f.9} with $M = 4$ and $\epsilon = 0.14$.  
The color of a node is based on its degree as indicated in the figure.} 
\label{f.11}
\end{figure}

\begin{figure}
\begin{center}
\includegraphics*[width=16cm]{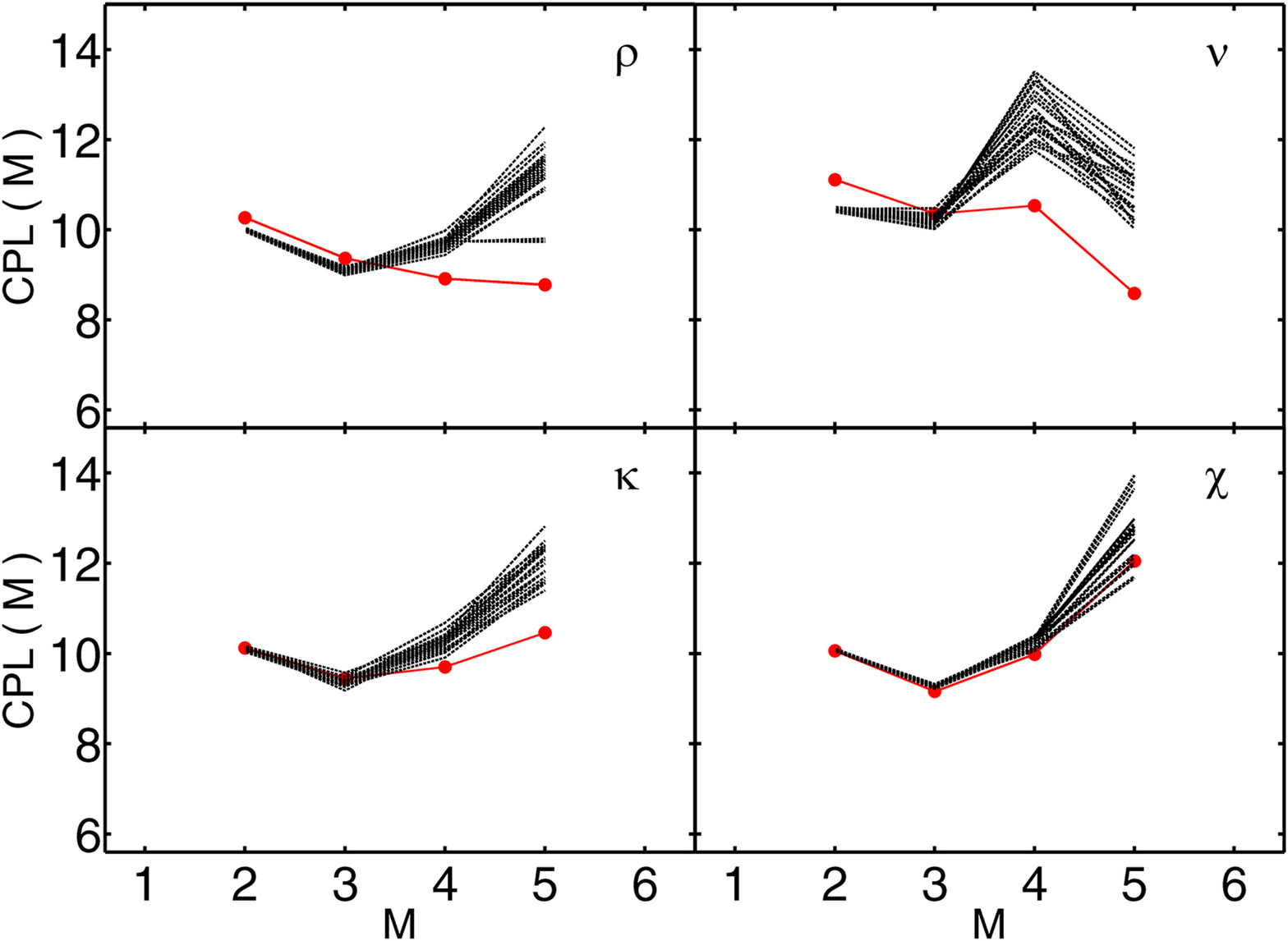}
\end{center}
\caption{Results of surrogate analysis of the four representative light curves from the 
black hole system with CPL as the discriminating statistic. Apart from the $\chi$ state, the other 
states show deviations from stochastic behavior and the null hypothesis can be rejected for them.} 
\label{f.12}
\end{figure}

Representative light curves from $4$ different classes are shown in Fig.~\ref{f.9}.  
The black hole system appears to 
flip from one class to another randomly in time and it is obvious that the light curves from each class 
are different even visually. The complete light curves have been presented in \cite {kph1}, where all 
the light curves have been analyzed using $D_2$ and $K_2$ and classified based on a dynamical 
perspective and the nature of the noise content. This classification is mostly confirmed by a 
recurrence plot analysis \cite {sukov} and very recently by a machine learning software 
analysis \cite {hupp}, where the authors developed a set of automated schemes based on supervised 
machine learning tools to efficiently classify the entire data set in terms of chaotic and 
stochastic processes.  Here we check whether the analysis based on 
network measures can provide more accurate information on the temporal behavior of these light curves.

Before constructing the RN from the light curves we show that our approach for the threshold selection 
can be applied for the real 
world data as well. In Fig.~\ref{f.10} (top panel), we show the RNs constructed from a representative 
light curve ($\nu$) for $M = 3$ with 3 values of $\epsilon$, namely, $0.06$, $0.10$ and $0.15$. In the 
bottom panel, the degree distributions of these RNs are also given. It is clear that $0.1$ can be taken 
as a reasonable choice of recurrence threshold. Thus the scheme works well for the real data as well.

The RNs constructed from the $4$ light curves shown in Fig.~\ref{f.9} are given in Fig.~\ref{f.11}. 
All the networks 
appear different from each other and the network for the $\chi$ state is similar  
to that from a white noise. We perform the RN analysis on light curves and their surrogates from all the 
$12$ classes taking $6$ different light curves for each class with the CPL as the quantifying measure. 
Results for the above $4$ states are shown in 
Fig.~\ref{f.12}. As expected, the $\chi$ state behaves identical to the white noise while the other three 
show deviations from purely stochastic behavior. When the average value of $nmsd$ for each class is 
compared with the threshold value for rejecting the null hypothesis as given above, we find that the 
null hypothesis can be rejected in all but $4$ states, namely, $\delta$, $\phi$, $\gamma$ and $\chi$,  
with the average $nmsd > 5.0$ in all the other states. Note that in Fig.~\ref{f.12}, the surrogates 
clearly deviate from data for $M \geq 4$ except for the $\chi$ state which is compatible with noise. 
We have found that the data and the surrogates deviate beyond a certain $M$ value for all the $8$ 
states for which the null hypothesis can be rejected.   

\begin{figure}
\begin{center}
\includegraphics*[width=12cm]{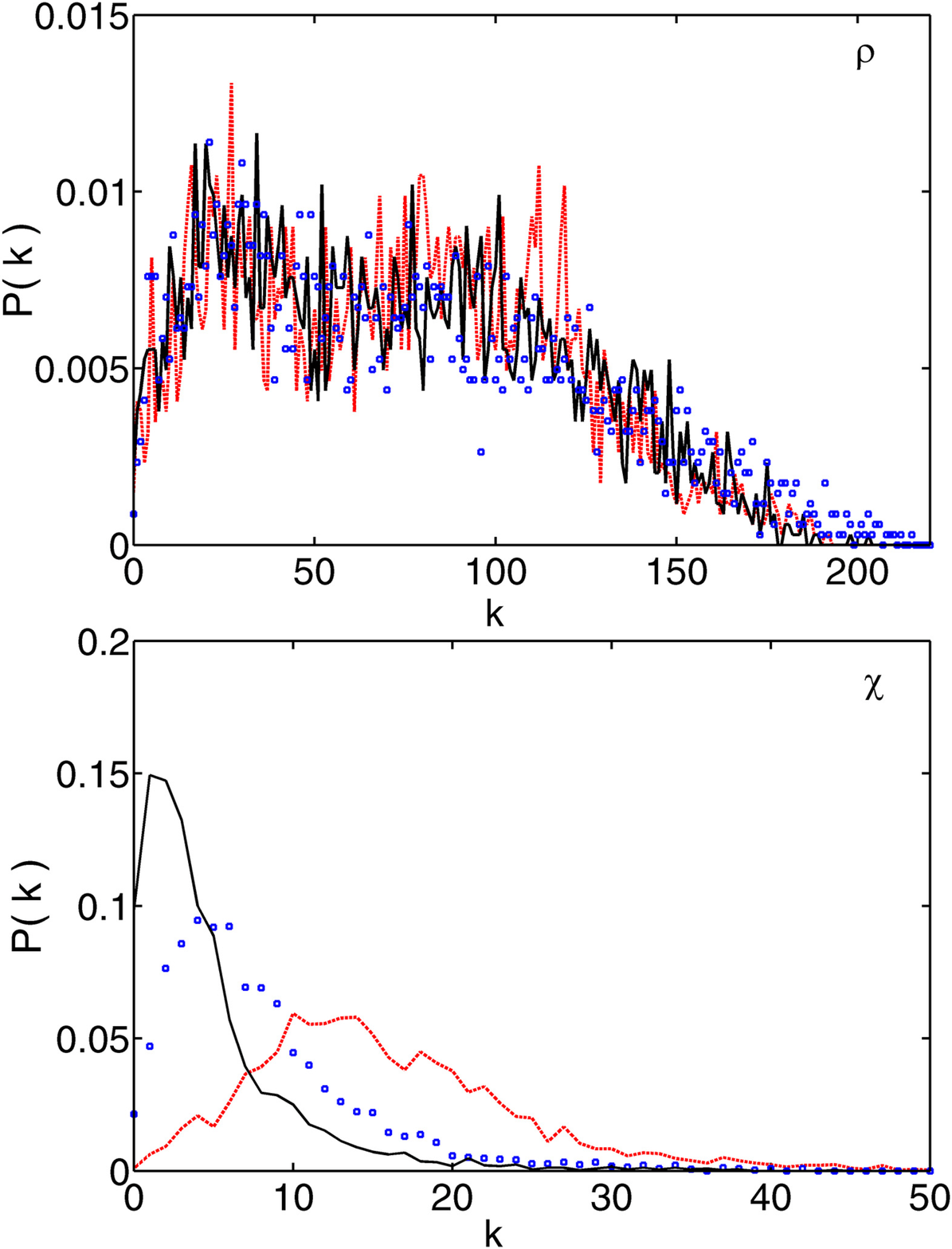}
\end{center}
\caption{Degree distribution of the RNs constructed from two representative GRS light 
curves with $M = 3$ (red dotted line), $M = 4$ (blue open circles) and $M = 5$ (black solid line). Note 
that the three degree distributions converge approximately for the $\rho$ state while they keep on 
shifting towards the left as $M$ increases for the $\chi$ state, which is the typical behavior of 
white noise.} 
\label{f.13}
\end{figure}

\begin{figure}
\begin{center}
\includegraphics*[width=12cm]{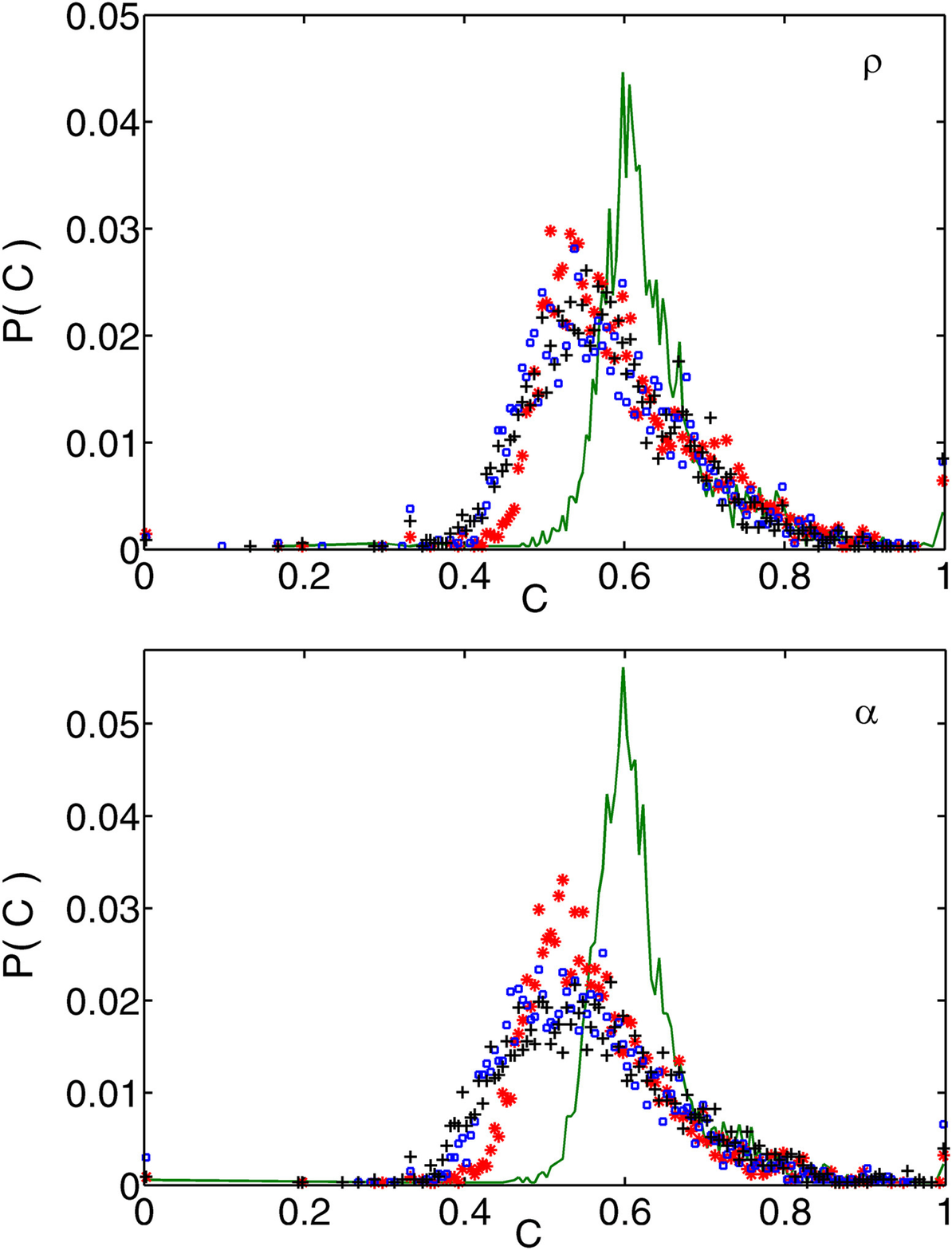}
\end{center}
\caption{Probability distributions of the local clustering coefficient $C$ for RNs constructed from 
the light curves of two GRS states for $M = 2$ (green solid line), $M = 3$ (red star points), 
$M = 4$ (blue open squares) and $M = 5$ (black crosses). In both cases, the distributions tend to 
converge for $M \geq 4$, indicating the dimensionality of both systems as $4$.} 
\label{f.14}
\end{figure}

We now compute the degree distribution of the RNs from all the light 
curves and find that except for the four states found to be compatible with noise in the RN analysis, 
all the other states show approximate convergence of the degree distribution with $M$. 
More importantly, the value of $M$ beyond which the convergence occurs is consistent with the 
$M$ value at which the data and the surrogates start deviating. The degree distributions 
for $2$ representative classes are shown in Fig.~\ref{f.13}. In three states out of $8$, the 
convergence occurs at $M = 3$ and in the remaining five, at $M = 4$. To get a convergence measure 
with $M$, we now compute the probability distribution $P(C)$ for all the states changing $M$ 
from $2$ to $6$. The distributions for two GRS states are shown in Fig.~\ref{f.14} and the 
K-L measures computed from the distributions as above, are shown in Table 1 for the complete 
range of $M$ values for all the $12$ states. The K-L measure does not show convergence with $M$ 
for the $4$ states for which the null hypothesis is rejected, indicating high dimensionality. 
The remaining $8$ states show convergence either at $M = 3$ or $M = 4$, as indicated in the last 
column of the table.

\begin{figure}
\begin{center}
\includegraphics*[width=12cm]{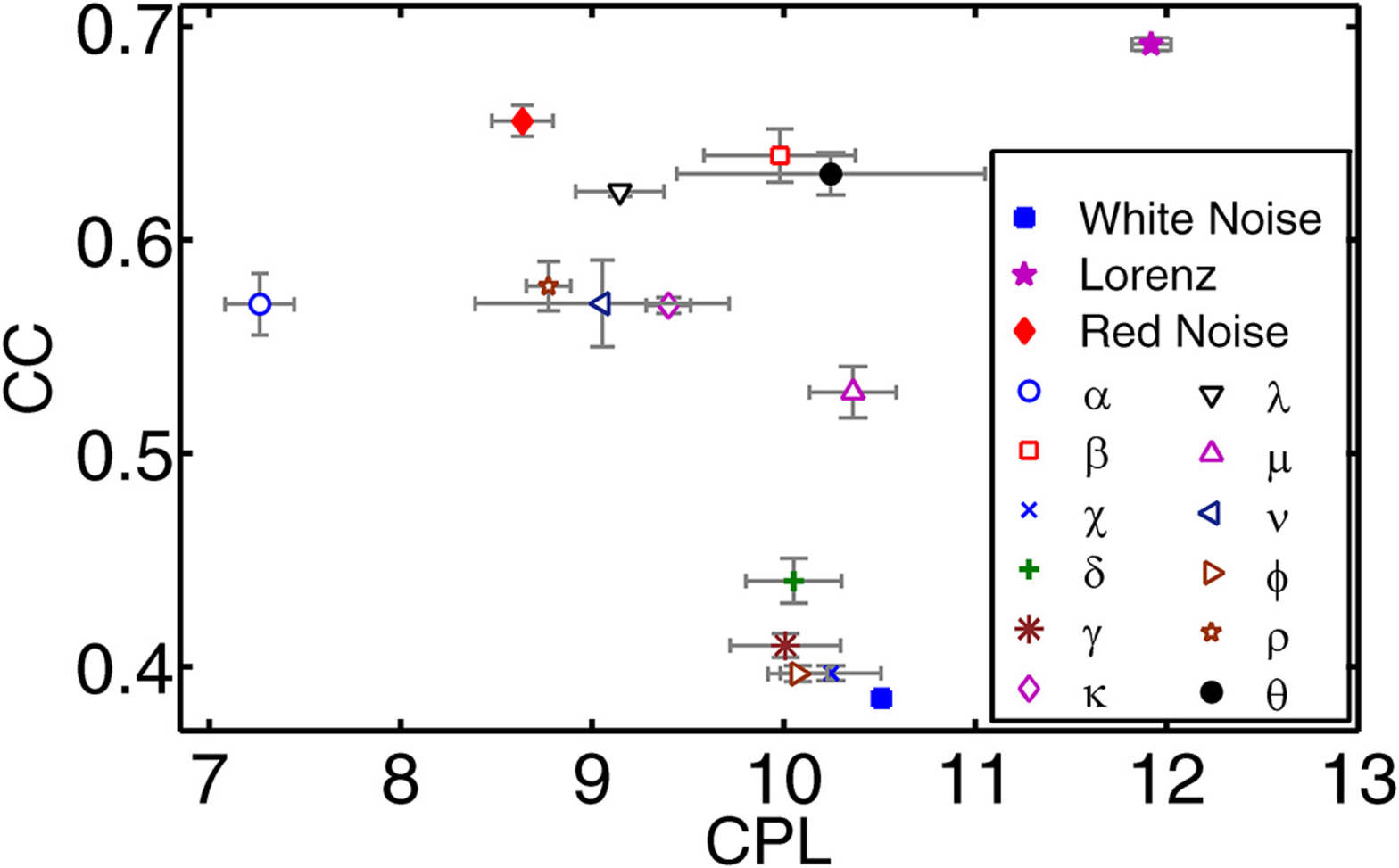}
\end{center}
\caption{Combined CPL-CC plot for all the GRS light curves along with the values for the Lorenz 
attractor, white noise and red noise for comparison. The error bar for the GRS states represents the 
variation on the value over the 6 light curves while that for synthetic data is the variation 
for 10 different simulations.} 
\label{f.15}
\end{figure}

Finally, we compute the global clustering coefficient CC for the RN using Eq. (2) for all the $12$ 
states taking $M = 4$ and present a combined CPL-CC plot. This is shown in Fig.~\ref{f.15}. To get 
a comparison with a genuine chaotic system and noise, we add the values for the RN from the 
Lorenz attractor, white noise and red noise. Note that the positions of the $4$ states for which the 
null hypothesis cannot be rejected are very close to white noise while of the remaining $8$ states,  
$7$ are clustered equally away from white and colored noise. Interestingly, one state $\alpha$, appears 
isolated with the CPL value very low compared to all other states. 
 
Combining the results from surrogate analysis, the K-L convergence measure and CPL-CC plot, we 
can arrive at the following conclusions:

a) The four states $\delta$, $\phi$, $\gamma$ and $\chi$ are consistent with a linear-stochastic process and 
null hypothesis cannot be rejected for them. They show properties very similar to white noise or 
$1/f$ noise (which we are unable to distinguish here). Hence we consider these $4$ states to be 
dominated by either white noise or $1/f$ noise. 

b) The remaining eight states appear to deviate from stochastic behavior with the 
underlying system having a finite dimensionality $M$ and are probably contaminated by some form of 
colored noise.       

Note that by null hypothesis, we are able to reject a specific type of stochastic process, but it does not 
enable us to accept any other alternative. In our computations, we have used only one specific type of 
colored noise for detailed analysis, namely, the red noise. There are other candidates in the category 
of colored noise and also other different possible stochastic processes that have not been tested. 
Hence it is difficult to derive any further conclusions other than those given above from our numerical 
results on the nature of the light curves. However, by closely analyzing the values of $nmsd$ of the states 
for which the null hypothesis is rejected, we try to get some further information regarding 
the nature of their temporal behavior. 

We find that the $nmsd(CPL)$ of the $8$ states for which linear stochastic behavior can be rejected 
fall into two different ranges. For five states 
$(\theta, \alpha, \beta, \lambda, \mu)$, we find the value of $nmsd$ in the range $5 < nmsd < 8$ and for 
the other three states $(\rho, \nu, \kappa)$, the values are $> 9$. Since moderate contamination of 
colored noise tends to keep $nmsd$ high and also tends to keep $M$ value saturated, we 
conjecture that the former $5$ states are \emph {colored noise dominated}. The latter three states 
can be considered to be potential candidates to search for \emph {deterministic nonlinearity}. 
Note that the results from the present analysis are mostly in agreement with the previous analysis 
using $D_2$ and $K_2$ \cite {kph1} where also the null hypothesis has been rejected in the same $8$ 
states. However, the colored noise contamination has become more evident in a couple of more states 
in the present analysis. 

\section{Discussion and conclusion}
The main objectives of the entire analysis undertaken here have been threefold:

i) To test the efficiency of RN measures as discriminating statistic for hypothesis testing using 
surrogate data

ii) To check whether these measures are effective if the data involves both white noise and colored 
noise which are very common in real world systems

iii) To illustrate the possibility of a network based approach to find the dimensionality of the 
underlying system if the time series deviates from a purely stochastic process.

We explored three particular network measures CPL, CC and the  distribution of $C$ in this analysis, with the 
first two used as discriminating measures. We applied a specific hypothesis that the data are derived from 
a linear stochastic process and then attempted to reject it using IAAFT surrogates. The time series from 
the standard Lorenz attractor added with different amounts of white noise, $1/f$ noise and red noise 
have been used to test the results of the analysis. We then found that CC is not a good discriminating measure 
if the data involves colored noise whereas CPL is effective in the presence of both white and 
colored noise. Hence we used only CPL as discriminating measure in the subsequent analysis of real data, 
the light curves from $12$ spectroscopic classes of the black hole system GRS 1915+105. 

The real advantage of using network measures is that they can be accurately computed from a lower number of  
nodes in the network, compared to the number of data points required for computing $D_2$ and $K_2$. 
One novel aspect of the present analysis is the result that the network approach can be 
combined with a new measure derived from the probability distribution of the clustering coefficient 
to get information regarding the dimension of the underlying system in the time series. Using this measure 
we are able to identify the dimension of  the light curves from all the temporal states for which the 
null hypothesis can be rejected. Thus we find that a network approach with CPL as discriminating 
measure and the convergence measure based on $P(C)$ is better suited to study the temporal properties 
of time series from the real world. 

Though our main motivation in the present analysis was to study application of network measures as tools 
to analyze real data, we think it is appropriate to mention what new information we have gained 
regarding the nature of light curves with 
this analysis. The present study mostly supports the previous one with null hypothesis rejected for 
the same states. One additional information of the present approach is the possibility of deriving the 
dimension also from a time series with a limited number of data points and a better information regarding 
colored noise contamination in various states.       
The astrophysical reason of why very few black hole systems show different spectroscopic classes 
(with qualitatively different temporal behavior) and random flippings between these classes, is still an 
open question and an active field of research.  The result that a few states may be deterministic and nonlinear 
does not in any way imply that there is evidence for \emph {chaos} in the system. It only implies that the 
accretion process responsible for the generation of the light curves in these states may have some 
underlying dynamics which can somehow be represented by a system  of coupled nonlinear differential equations. 
Chaos, of course, is a more exciting prospect, but warrants more specific criteria on the part of the 
system apart from nonlinearity. We are of the opinion that  
detecting chaos from limited real world data immersed in 
noise using any of the time series methods is extremely difficult (if at all possible). In the 
present context, it requires the development of, at least, a truncated model of the 
accretion process in which critical changes in one or two control parameters can bring about qualitative 
changes in the nature of the light curves. This is, by far, a very challenging task.   

{\bf Acknowledgements} 

We thank one of the anonymous referees for a thorough review of the manuscript and suggesting 
many changes to improve the presentation. 
RJ and KPH acknowledge the financial support from Science and Engineering Research Board (SERB), 
Govt. of India in the form of a Research Project No SR/S2/HEP-27/2012. 
KPH  acknowledges the computing facilities in IUCAA, Pune.  

For graphical representation of networks, we use the GEPHI software: \\  
\emph {(https://gephi.org/)}.

\bibliographystyle{elsarticle-num}

\end{document}